\documentclass[aip,
 amsmath,amssymb,
 preprint,%
]{revtex4-2}

\usepackage{graphicx}
\usepackage{dcolumn}
\usepackage{bm}

\usepackage[utf8]{inputenc}
\usepackage[T1]{fontenc}
\usepackage{mathptmx}

\begin{document}

\preprint{AIP/123-QED}

\title{Giant interfacial induced Dzyaloshinskii-Moriya interaction in polycrystalline Co/Pt multilayers}


\author{D.A. Tatarskiy}
    \email{tatarsky@ipmras.ru.}
    \affiliation{Institute for physics of microstructure of RAS, Akademicheskaya st., 7, Afonino, 603087, Russia}
    \affiliation{Lobachevsky University, Gagarin ave., 23, Nizhniy Novgorod, 603950, Russia}

\author{N.S. Gusev}
\author{S.A. Gusev}
    \affiliation{Institute for physics of microstructure of RAS, Akademicheskaya st., 7, Afonino, 603087, Russia}


\date{\today}

\begin{abstract}
Multilayers of heavy metals and ferromagnets are the basis for the formation of various structures with inhomogeneous chiral magnetic distributions. The most informative method for studying magnetic states in such structures with a high spatial resolution is Lorentz transmission electron microscopy (L-TEM).  Here, we report on the observation of  chiral N\'eel domain walls by means of L-TEM in Co/Pt multilayers with perpendicular magnetic anisotropy. We explain the existence of chiral N\'eel internal structure due to the overcritical, giant interfacial induced Dzyaloshinskii–Moriya interaction. Our results are confirmed by micromagnetic modeling, which allows taking into account the effect of local inhomogeneity of the magnetic anisotropy of polycrystalline films on formation of spin textures.
\end{abstract}

\maketitle

\section{Introduction}
Noncollinear spin textures is one of the most interesting and intensively studied objects of the magnetism during last years due to their unusual topological and transport properties and their potential applications for data storage and information carriers in low-power nanoelectronics~\cite{Yu,Nagaosa}. Multilayer films, consisting of ferromagnetic (FM) and heavy metal (HM) layers, are the most popular systems for these magnetic applications because their magnetic properties can be fine-tuned by appropriate choice of materials and layer thicknesses, total number of layers and their crystal structure. In these multilayer films with interfacial induced Dzyaloshinskii-Moriya interaction~\cite{Dzyal,Moriya,Bogdanov,Anatomy} (iDMI), it is possible to stabilize nanoscale spin textures such as skyrmions at room temperatures~\cite{Boulle,ACSSksRT}. Thus, systems of thin metal films of nanometer thickness with iDMI, due to the relatively simple technology of their formation, seem to be the most promising structures as a basis for constructing logical~\cite{Fert,RaceTrack,LogicW} or neuromorphic~\cite{Neuro} magnetic devices.

It is also known that the presence of iDMI in multilayer metal films leads to topologically  homochiral 360 degree N\'eel domain walls~\cite{Walls01} in media with perpendicular magnetic anisotropy. Also N\'eel skyrmions and N\'eel walls are experimentally observed in some types of multilayers consisting of films of three different materials~\cite{TEM1,TEM3,TEM4,TEM5,TEM6,Walls02,Walls03,Walls04,Walls06,Wang201912098,Speelm,MFM}. In such structures, the magnetic layers are in contact with layers of heavy metals and oxides or with two different heavy metals. In this case, the value and sign of iDMI at the upper and lower boundaries of the ferromagnetic layer may have different values, and the total iDMI value is expected to be the greatest one, and this will improve the stability of skyrmions formation and reduce their size. In addition to such structures, there is some oblique evidence for the presence of iDMI in quasi-symmetric multilayers~\cite{TEM2,Walls05} of two types of metal layers Co/Pd. But these multilayers have platinum capping and underlayers, and they are not purely symmetrical from a material point of view. Obviously the iDMI is canceled in ideal system consisting of two materials. That is true for purely epitaxial interfaces~\cite{Epitaxy1}. Still the iDMI value can be tuned by material concentration variations in a top layer~\cite{Epitaxy1} or by growth conditions~\cite{Epitaxy2}. One should mention that observed iDMI value in epitaxial films is few times smaller than it should be from the first principle calculations even taking into account interface roughness~\cite{Rough}.

The most common method used in manufacture of multilayer metal structures is magnetron sputtering. With this manufacturing method, the metal films have a polycrystalline structure with imperfect interfaces due to the roughness of the boundary layers, different orientations of nanocrystals or the mixing of ferromagnet and heavy metal. These imperfections can introduce asymmetry into the structure of surfaces ``FM/HM'' and ``HM/FM'' of the multilayer, and therefore, the iDMI value in multilayers can differ at the ``top'' and ``bottom'' interfaces of the ferromagnetic film~\cite{abinitioInterlayer}. Thus, the presence of iDMI due to the structural asymmetric interfaces can be expected in magnetron-deposited polycrystalline multilayers, which consist of only two metals, for example, cobalt and platinum.

Another way to get iDMI in the multilayer films system is an interlayered mechanism due to spin-orbit coupling not on the interface but on interlayers of heavy metal~\cite{BergerInterlayer1,BergerInterlayer2}. Recently this mechanism of iDMI has been demonstrated in cobalt-palladium multilayer between platinum top and bottom layers~\cite{TEM2}. The most fascinating question wether one can get large enough iDMI in more homogeneous films, e.g. cobalt-platinum multilayer instead of cobalt-palladium.

There are several experimental methods which can be applied to study the distribution of magnetization in nanostructures and can provide information on the presence of the iDMI in thin film systems. The Brillouin light scattering~\cite{BLS4} and neutron scattering technique~\cite{Monchesky,B201} are used to study the Dzyaloshinskii-Moria interaction. However, these methods have a low spatial resolution, which does not allow to obtain local magnetization distributions arising in structures with iDMI. The most qualitative and quantitative method for studying inhomogeneous states in thin films with a high spatial resolution is Lorentz transmission electron microscopy (L-TEM). In this paper, we present the experimental results using the L-TEM method, which demonstrate the existence of iDMI in multilayer Co/Pt films. The micromagnetic modeling was used to show the effect of nanocrystalline structure of Co and Pt layers on magnetization distributions in multilayer films, which allowed us to explain the features of the experimentally observed Fresnel contrast.

\section{Samples preparation and methods}

The magnetic multilayer films with Ta(3.2)/Pt(5.5)/[Co(0.5)/Pt(1.0)]$_{\times 5}$/Pt(3.8) structure (where 5 denotes the layer repetition numbers, thickness in nm) are grown by DC magnetron sputtering on 50-nm-thick Si$_3$N$_4$ membrane windows. The films are fabricated in a high vacuum magnetron AJA 2200 system at the basic argon pressure of  $4 \times 10^{-3}$ Torr. The preliminary pumping of the working chamber is carried out to a vacuum of $10^{-5}$ Torr. The sputtering is carried out from separate Ta, Co, and Pt targets at room temperature. The deposition rates of the materials are 0.1 nm/s  for cobalt films and 0.3 nm/s for platinum. The thicknesses of the films are determined by the deposition time after preliminary calibration of the thicknesses of individual layers, measured by X-ray reflectometry.  The additional 2 nm platinum layer is deposited on the top periodic structure to avoid oxidation. Recently, we have experimentally shown that the thicknesses of the layers obtained by this technology are in good agreement with the calculated values by studying the cross-sections of similar structures with transmission electron microscopy~\cite{Tatarskiy}.

\begin{figure}
\begin{minipage}[h]{0.32\linewidth}
\center{\includegraphics[width=0.95\linewidth]{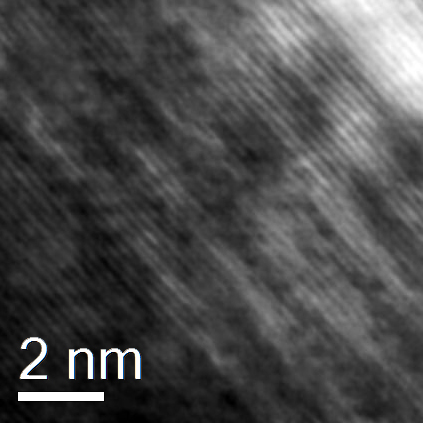} \\ (a)}
\end{minipage}
\hfill
\begin{minipage}[h]{0.32\linewidth}
\center{\includegraphics[width=0.95\linewidth]{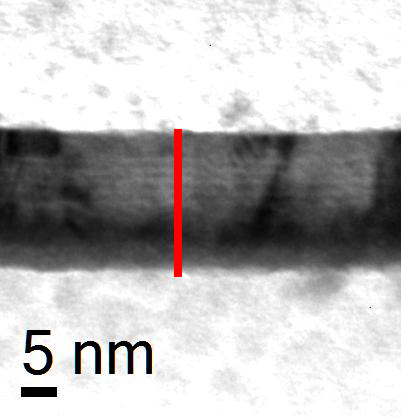} \\ (b)}
\end{minipage}
\hfill
\begin{minipage}[h]{0.32\linewidth}
\center{\includegraphics[width=0.95\linewidth]{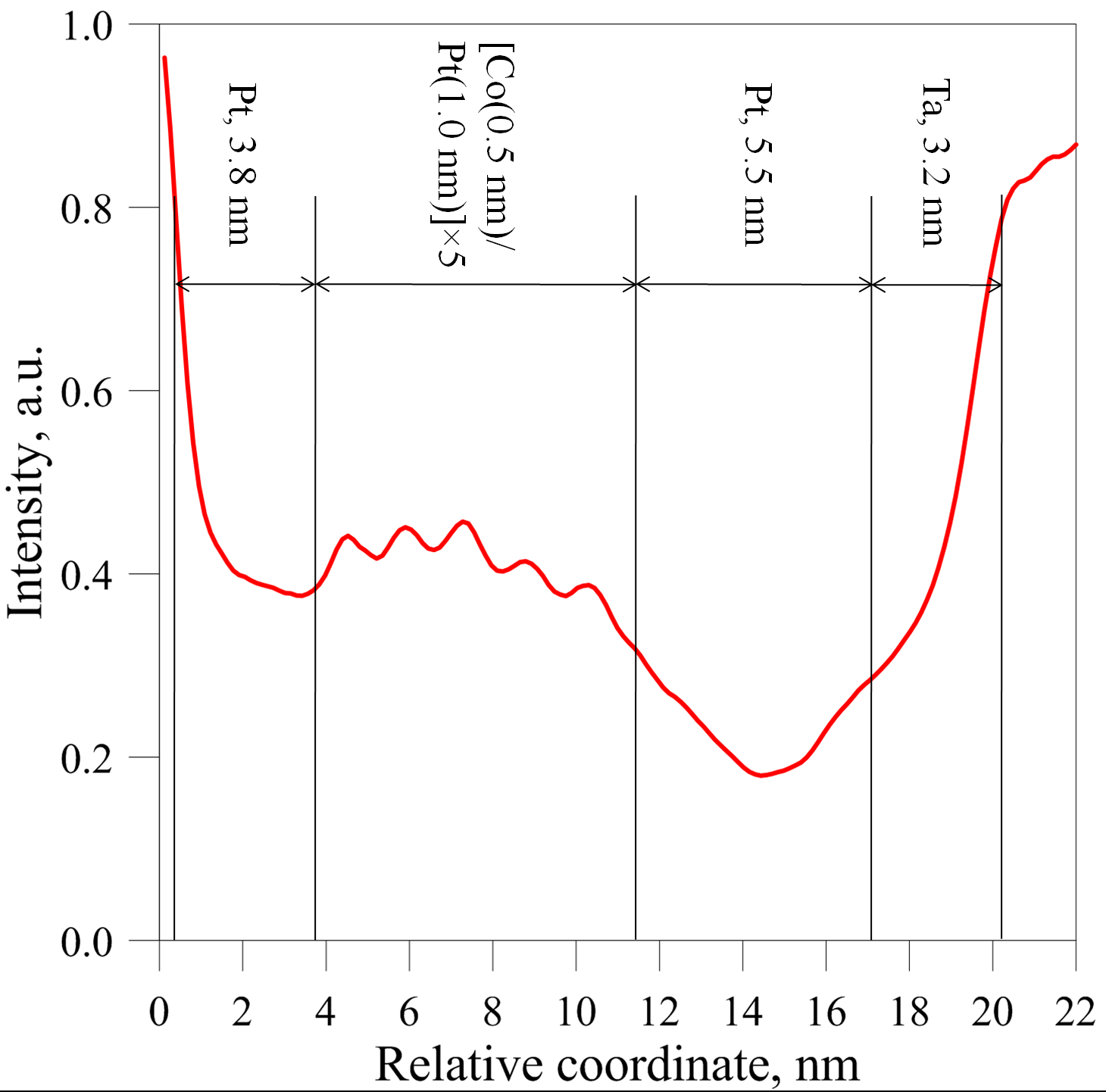} \\ (c)}
\end{minipage}
\caption{\label{Fig0} (a) -- High resolution micrograph of multilayer, the distance between atomic planes is $\sim 0.225$ nm; (b) -- Bright field overview of the co/Pt multilayer; (c) -- Average profile of intensity, extracted from (b) along red line.}
\end{figure}

Magnetic properties of the Co/Pt samples are studied by magneto-optical Kerr effect (MOKE) measurements using custom-made magneto-optical stand with 632 nm wavelength He-Ne laser. Transmission electron microscopy performed by LIBRA200 MC (Carl Zeiss) instrument operating at 200 kV is used to study the microstructure of the films. We study the cross-sectional structure of multilayer films and analyzed in-plane micrographs. TEM lamella preparation is performed by FIB-SEM crossbeam station Auriga Laser (Carl Zeiss). The fabricated cross-sections are also polished using a low-energy ($\sim$ 300 eV) Ar$^+$ source to reduce the thickness of the damaged layers. The L-TEM measurements are performed by C$_s$--corrected TITAN 80-300 transmission electron microscope (FEI) operated at 80 kV. In these experiments, it is possible \textit{in situ} to observe changes in the Fresnel contrast caused by an external magnetic field, which is created at the location of the sample due to weak excitation of the microscope objective lens and tilting the sample to re-magnetize it. The maximum value of this magnetic field in our experiments was 35 mT.

We use MuMax$^3$ GPU-accelerated micromagnetic software~\cite{Mumax3} to simulate the distributions of magnetization in multilayer structures consisting of polycrystalline films. Polycrystalline grains in  MuMax$^3$ are simulated by the Voronoi tessellation method~\cite{Voronoi}. We suppose each grain has its own value and direction of uniaxial anisotropy and iDMI strength. The film as a whole has an average nonzero anisotropy along its normal. For the sake of simplicity we use model of random uniaxial anisotropy in each crystalline grain even if the crystallographic structure can be cubic or hexagonal as for typical cobalt, platinum and their alloys. Other micromagnetic parameters (the saturation magnetization, the exchange stiffness) are homogeneous over the film and their values are the same as in our previous papers~\cite{Bubbles,Materials}. The distributions of magnetization in the structures obtained by numerical calculations are used to simulate images with magnetic contrast in the transmission electron microscope.  To simulate L-TEM images, a specially developed original Python script for the Gatan Microscopy Suite Software\textsuperscript{\textregistered} 3.4.1 is used~\cite{PyFresnel}. It is known that the value of the contrast $\Delta I$ in Fresnel mode is proportional to the projection of magnetization curl on the optical axis~\cite{McVitieSim1,McVitieSim2} of an electron microscope and defocus value $\Delta z$
\begin{equation}\label{FresnelContrast}
\Delta I \left( \vec{\rho} \right) \sim \Delta z \left(\vec{n} \cdot \left[ \vec{\nabla} \times \vec{M} \left( \vec{r} \right) \right] \right).
\end{equation}
where $\vec{\rho}$ is a vector in the detector plane and $\vec{r}$ is a vector in the plane of the film. In general case the sample is tilted and the film's and detector's plane do not coincides. The total value of the intensity (taking into account the material contrast $I_0(\vec{\rho})$ is an inverse Fourier transform (IFFT) of the convolution of the Fourier image of~(\ref{FresnelContrast}) with the phase transfer function~\cite{deGraefSim}
\begin{equation}\label{DetectorContrast}
I_{d} \left( \vec{\rho} \right) \sim \left| \textrm{IFFT} \left[ \int \left( I_0 \left( \vec{q_\bot} \right) + \Delta I \left(\vec{q_\bot}\right) \right) \exp \left( \pi \lambda \Delta z \left| \vec{q}_\bot \right|^2 \right) \exp \left( -\frac{\left(\pi \theta \Delta z \right)^2 \left| \vec{q}_\bot \right|^2}{ln 2} \right) d\vec{q_\bot} \right] \right|,
\end{equation}
where $\lambda$ is electron wavelength, $\theta$ is beam divergence and $\vec{q}_\bot$ is a reciprocal lattice vector in the detector's plane. The Fresnel contrast simulations are done taking to account TEM transfer function for 80 keV electron energy, defocus 2 mm and beam divergence $\theta = 0.09$ mrad.  Then these calculated images are compared with the LTEM images obtained experimentally.

\begin{figure}
\begin{minipage}[h]{0.32\linewidth}
\center{\includegraphics[width=0.95\linewidth]{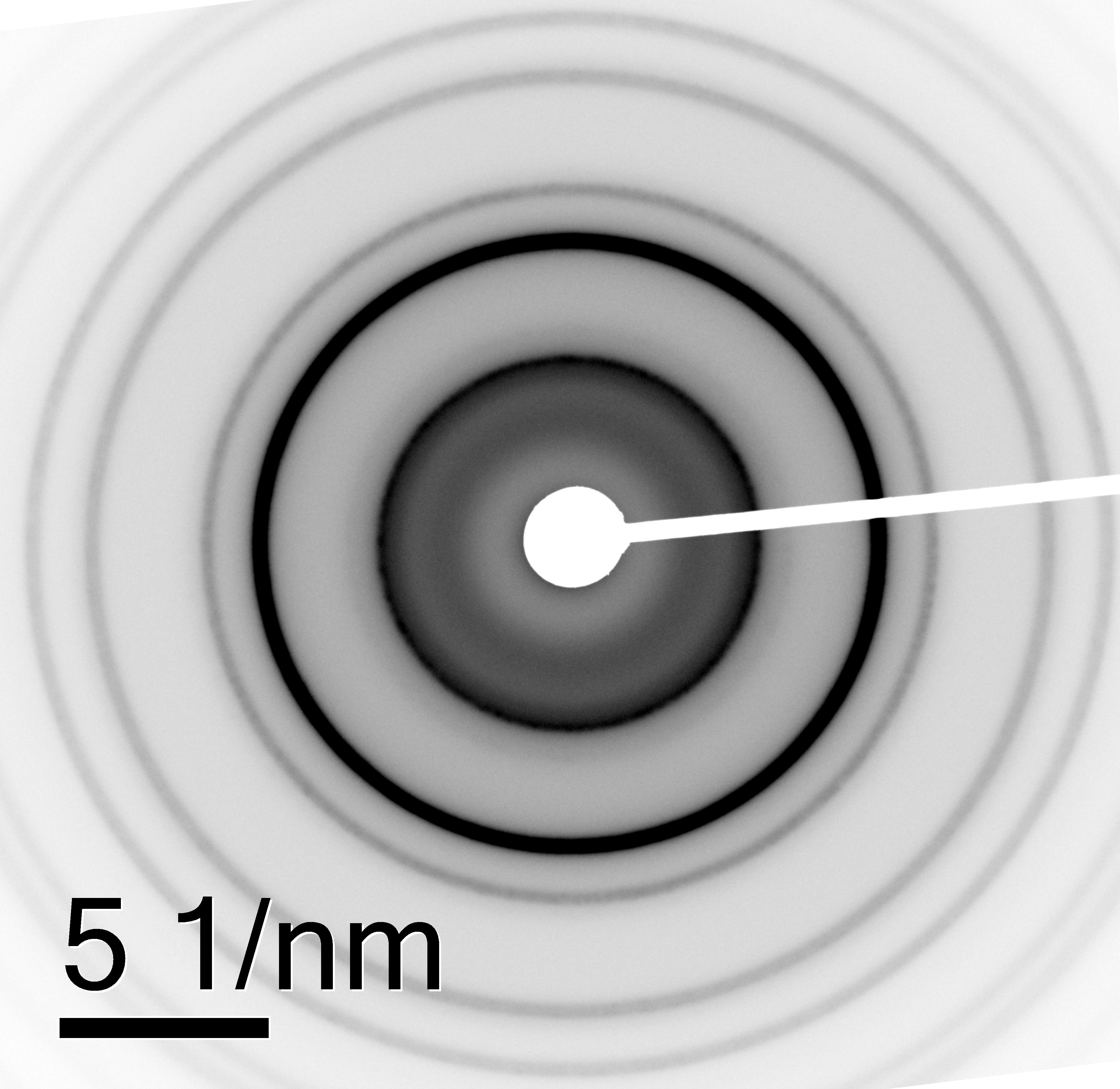} \\ (a)}
\end{minipage}
\hfill
\begin{minipage}[h]{0.32\linewidth}
\center{\includegraphics[width=0.95\linewidth]{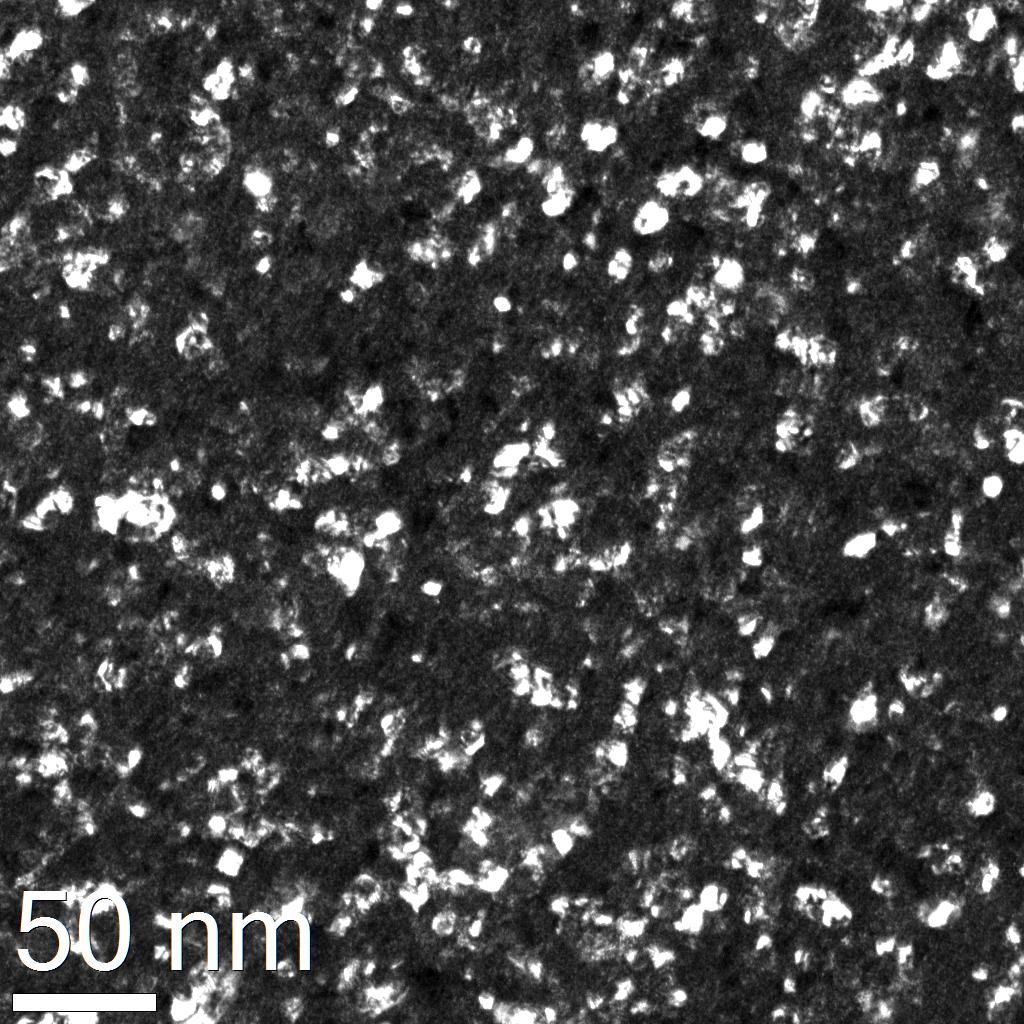} \\ (b)}
\end{minipage}
\hfill
\begin{minipage}[h]{0.32\linewidth}
\center{\includegraphics[width=0.95\linewidth]{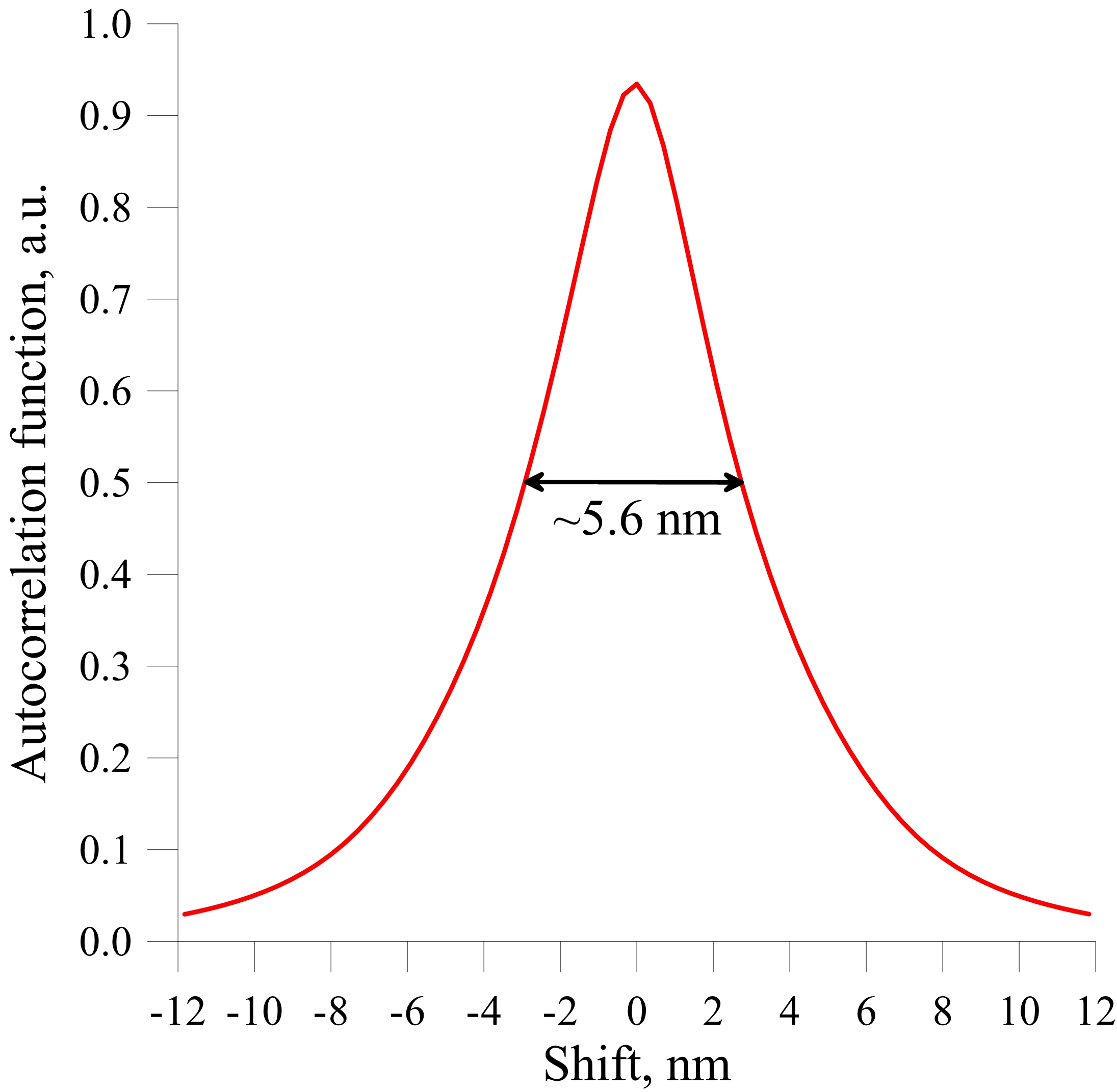} \\ (c)}
\end{minipage}
\caption{\label{Fig1} (a) -- Microdiffraction pattern; (b) -- the dark field micrograph of Co/Pt multilayer; (c) -- autocorrelation function of dark field micrograph.}
\end{figure}

\section{Results and discussions}
TEM images are obtained for the overall characterization of the microstructures of Co/Pt multilayers. The high-resolution TEM image (HRTEM) of the Co/Pt cross-section clearly shows the crystal lattice of cobalt and platinum layers with an interplanar spacing of 0.225 nm (Fig.~\ref{Fig0}a)), which can best correspond to the (111) plane of the $Co_{1+x}Pt_{3-x}$ solid solution with Pm3m space group of symmetry lattice. However, the Bright Field (BF) image of a cross-section of a multilayer film shows the contrast between the Co and Pt layers (Fig.~\ref{Fig0}b) which indicates the layered structure of the sample.  The average value of the Co layer thickness measured from the averaged profiles of micrograph contrast  (Fig.~\ref{Fig0}c) is 0.5 nm, and the thickness of the Pt layer is 1.0 nm ,which is in good agreement with the values that were set by the time of the magnetron deposition of films. Analysis of TEM in-plane images of the film provides additional information on the microstructure of the Co/Pt sample. The Selected Area Diffraction (SAD) patterns of a film corresponds to a cubic lattice (Fig.~\ref{Fig1}a) with lattice parameter $a = 0.390 \pm 0.008$ nm. This value is between the parameters of platinum lattice $a_{Pt} = 0.392$~nm and cubic lattice of solid solution $CoPt_3$ $a_{CoPt} = 0.384$~nm, which shows the presence of elastic stresses in the multilayered structure. A typical Dark Field (DF) micrograph of Co/Pt polycrystalline multilayer and its autocorrelation function is shown on Fig.~\ref{Fig1}b,c. The full width at half maximum of autocorrelation function is $\sim 5.6$ nm. This value can be taken to account for calculation of volume-weighted average grain size~\cite{ACF1,ACF2}, and is used by us in the simulations. Basing upon these measurements we fit the experimental MOKE hysteresis curve by micromagnetic simulations.

\begin{figure}
\center{\includegraphics[width=0.48\linewidth]{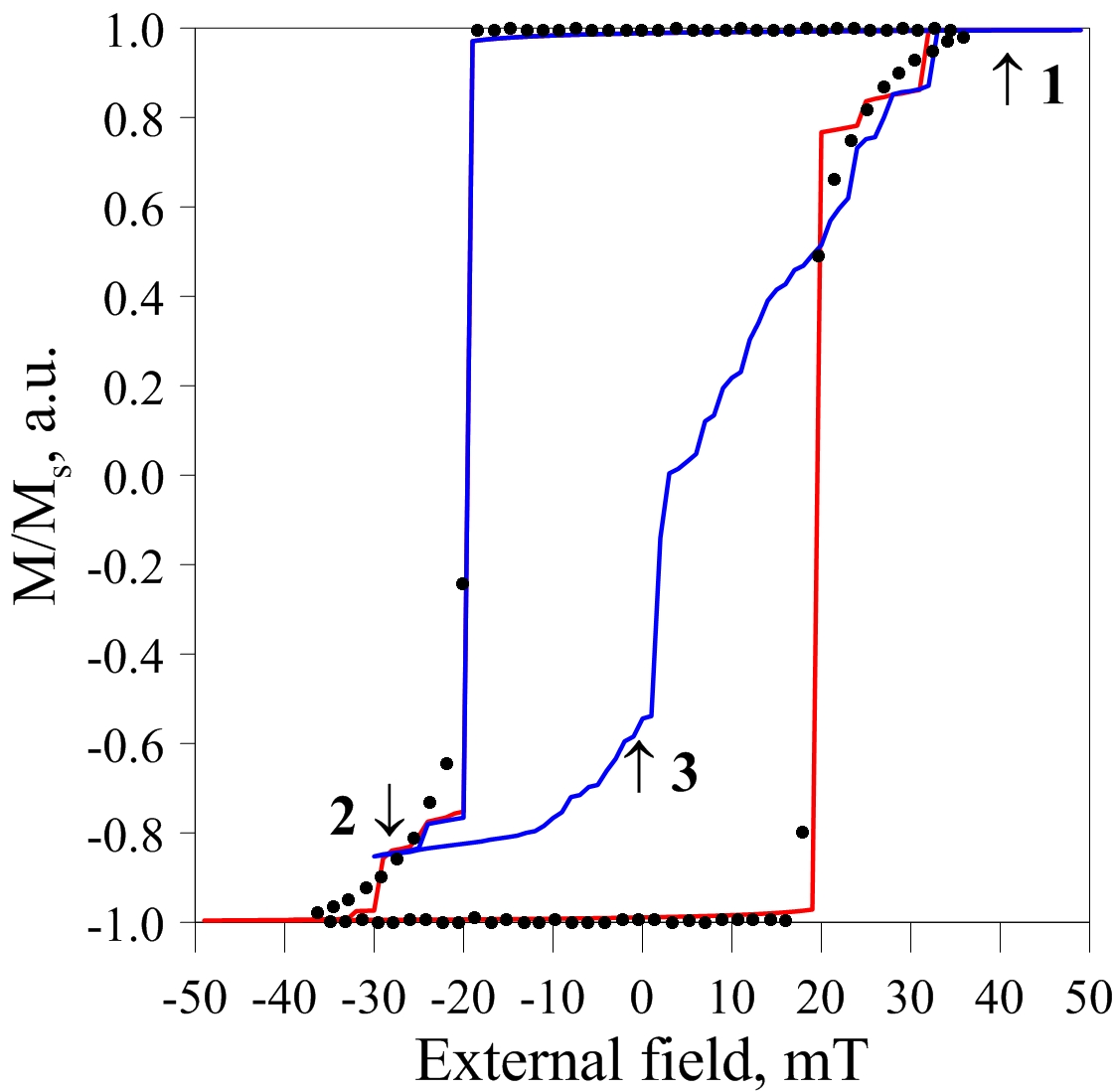}}
\caption{\label{Fig2} Circles -- magneto-optical measurements; red curve -- simulated hysteresis; blue curve -- minor hysteresis, numbers indicate where the Fresnel contrast is acquired through \textit{in situ} remagnetization.}
\end{figure}

We simulate the $256 \times 256 \times 7.5$~nm$^3$ part of the film with $0.5 \times 0.5 \times 7.5$~nm$^3$ unit cell.   A total count of crystallites was around $\sim 1800$, but only 256 different material parameters (variations of magnetic anisotropy value and its direction, iDMI value etc.) were taking into account due to the MuMax$^3$ software restrictions. In other worlds every 7 crystallites in different parts of simulation have equal material parameters. This approach demonstrates a good approximation of a real random polycrystalline film. From our previous experiments we suppose the values for magnetic parameters of material~\cite{Materials,Bubbles}.
\begin{figure}
\begin{minipage}[h]{0.32\linewidth}
\center{\includegraphics[width=0.95\linewidth]{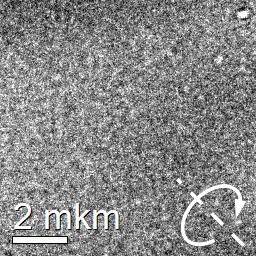} \\ (a)}
\end{minipage}
\hfill
\begin{minipage}[h]{0.32\linewidth}
\center{\includegraphics[width=0.95\linewidth]{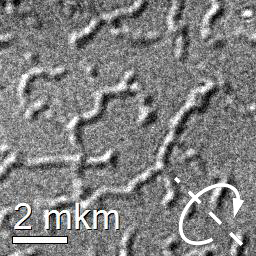} \\ (b)}
\end{minipage}
\hfill
\begin{minipage}[h]{0.32\linewidth}
\center{\includegraphics[width=0.95\linewidth]{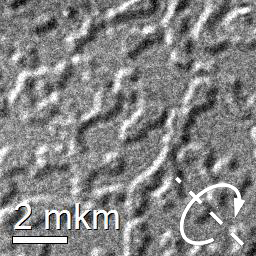} \\ (c)}
\end{minipage}
\caption{\label{TITAN} Fresnel contrast on micrographs. Figures are acquired at points 1,2,3 from hysteresis curve. The contrasts are obtained at 30 degrees tilt along denoted axis.}
\end{figure}
The difference is that in this work we took into account a nonzero value iDMI, which could have different values for different crystallites of the film. The iDMI value for each crystallite was set randomly, and the maximum deviation did not exceed $15\%$  from the average value. In this range of deviations from the mean value, the magnetic anisotropy value $K$ for each nanocrystal was also randomly set. The set of material parameters for modeling was chosen so as to ensure the best match between the simulation and experimental MOKE and L-TEM data. The best fit is achieved for magnetization $M_s = 200$ kA/m, exchange stiffness $J=7 \times 10^{-13}$ pJ/m, $K = 3.1 \times 10^4 \pm 15 \%$ J/m$^3$, wandering of anisotropy direction from film normal up to $8.5$ degrees, iDMI value $\left| D \right| = 60 \pm 15\%$ $\mu$J/m$^2$. The micromagnetic simulations demonstrates not only a good quantitative agreement with a magneto-optical measurements~(Fig.\ref{Fig2}) but a good agreement with obtained Fresnel contrasts during \textit{in situ} remagnetization of the sample (compare Figs.~\ref{TITAN} and~\ref{MicromagSim}).

\begin{figure}
\begin{minipage}[h]{0.24\linewidth}
\center{\includegraphics[width=0.95\linewidth]{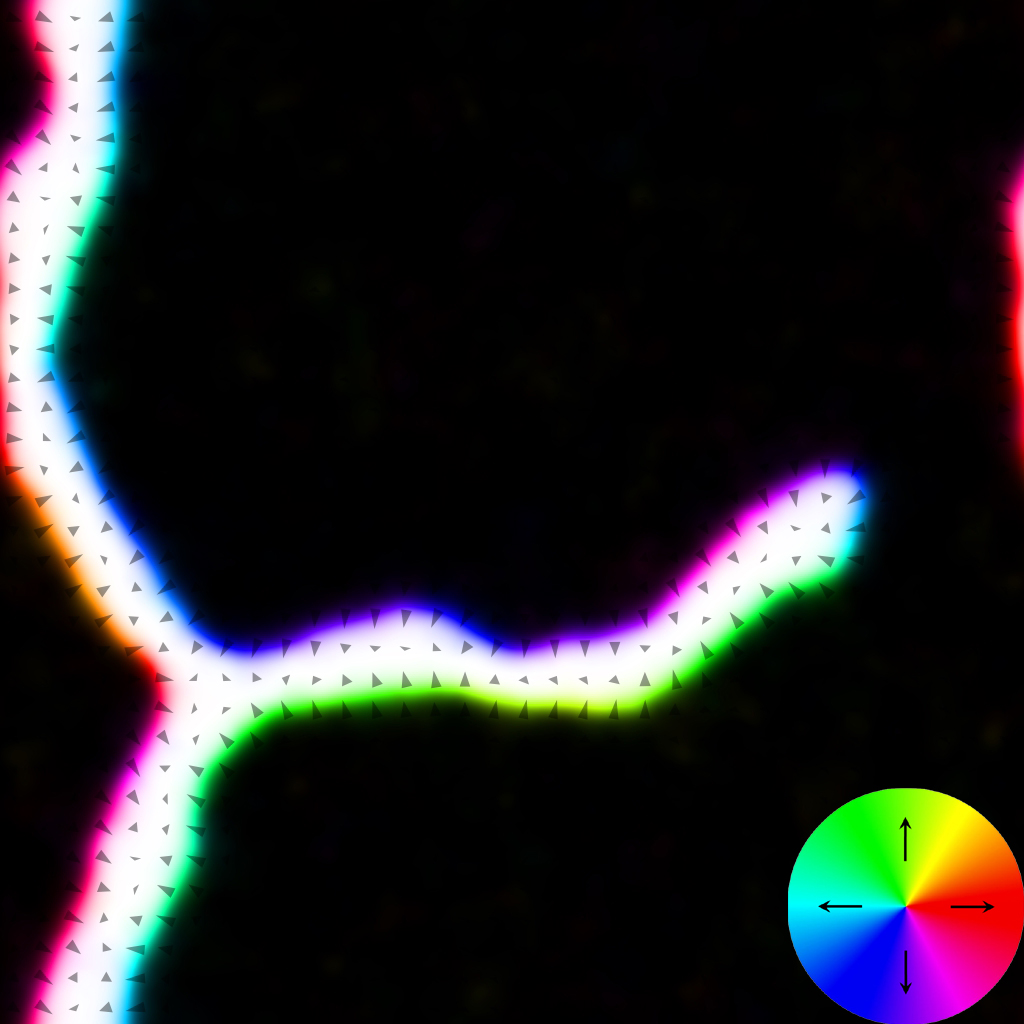} \\ (a)}
\end{minipage}
\hfill
\begin{minipage}[h]{0.24\linewidth}
\center{\includegraphics[width=0.95\linewidth]{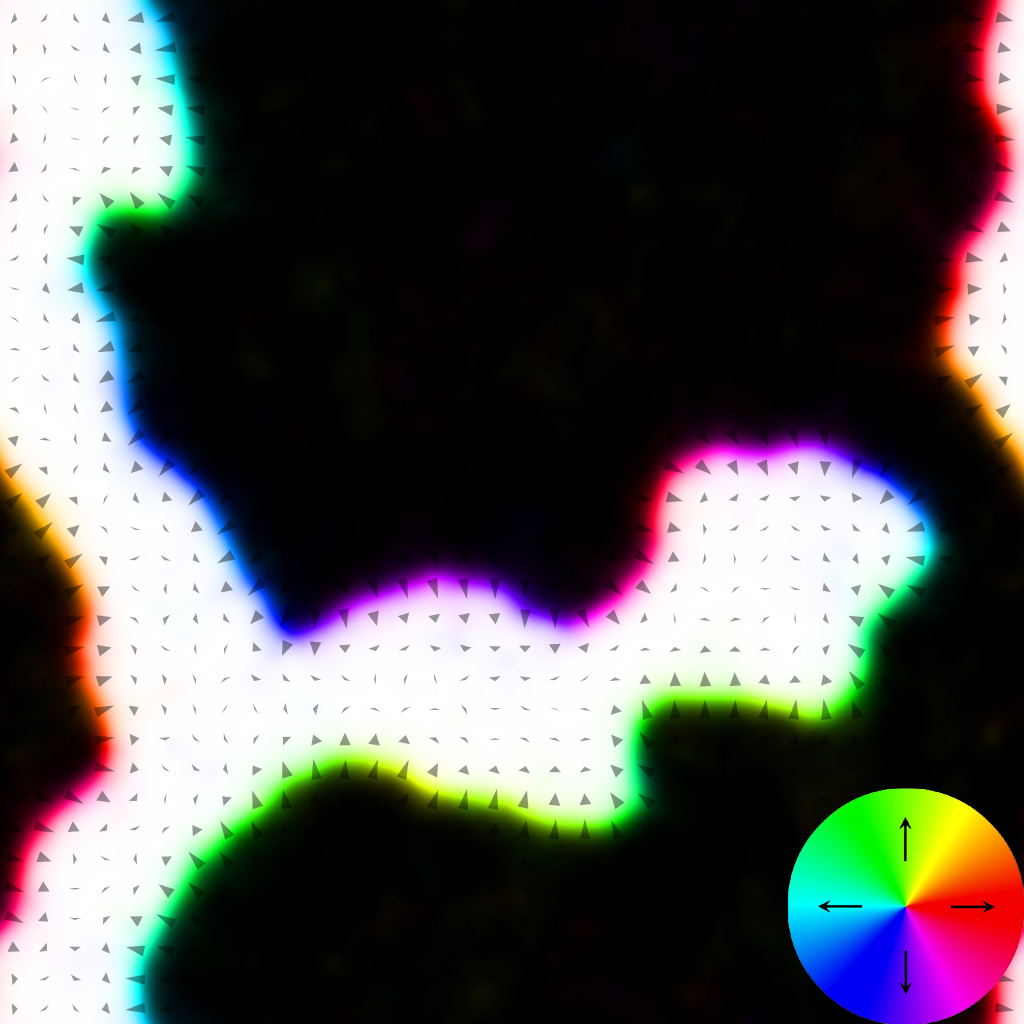} \\ (b)}
\end{minipage}
\hfill
\begin{minipage}[h]{0.24\linewidth}
\center{\includegraphics[width=0.95\linewidth]{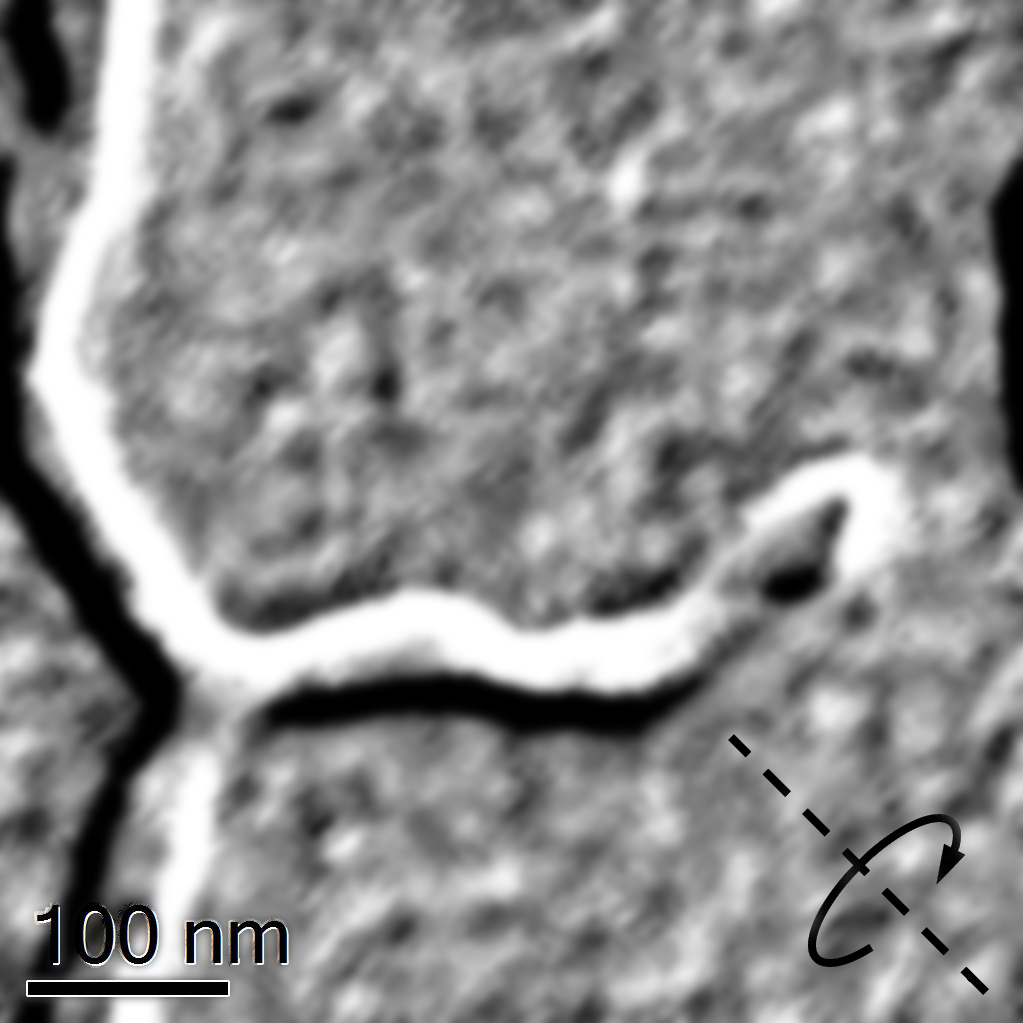} \\ (c)}
\end{minipage}
\hfill
\begin{minipage}[h]{0.24\linewidth}
\center{\includegraphics[width=0.95\linewidth]{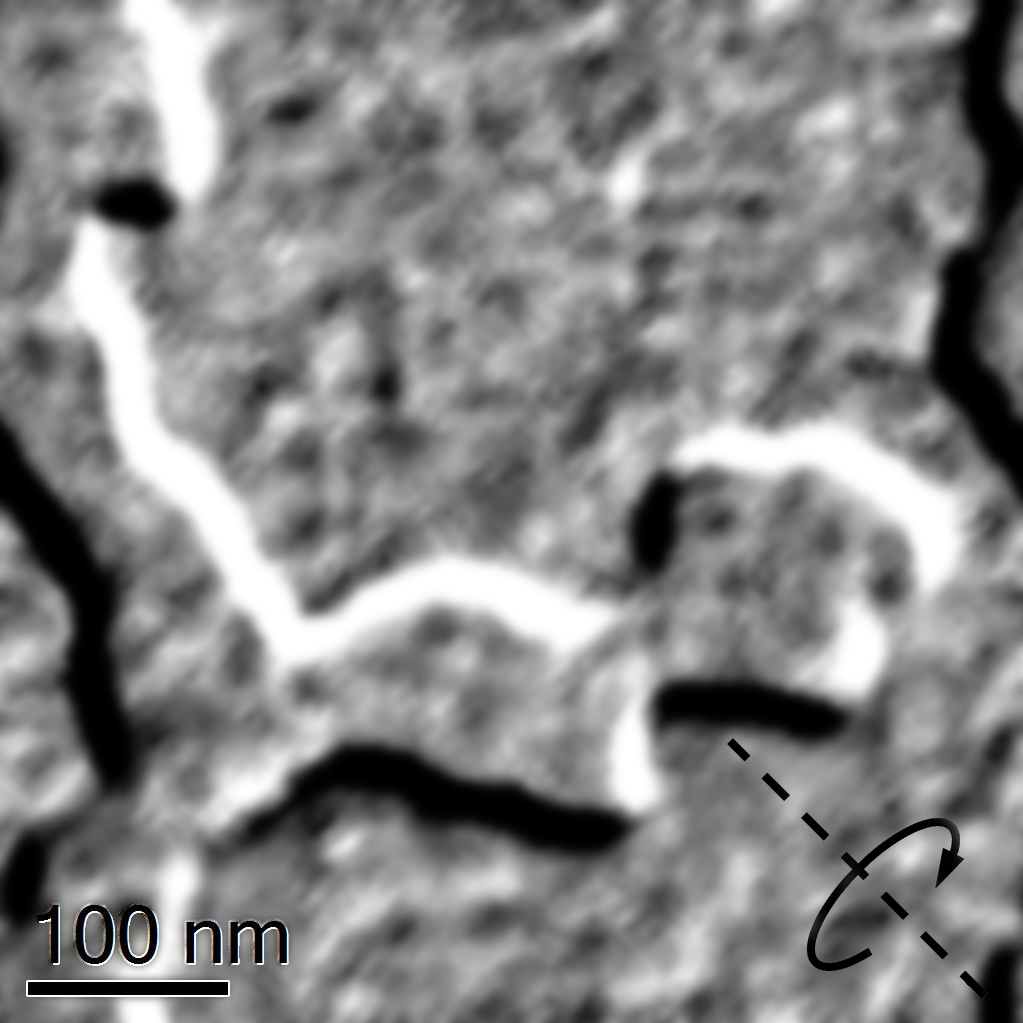} \\ (d)}
\end{minipage}
\caption{\label{MicromagSim} (a),(b) Simulated magnetization distributions of points 2 and 3 from hysteresis curve, rainbow color denote in-plane magnetization direction, black-white colors denote $M_z$ direction; (c),(d) -- simulated Fresnel contrast at 30 degree tilt along diagonal axis.}
\end{figure}

\begin{figure}
\begin{minipage}[h]{0.24\linewidth}
\center{\includegraphics[width=0.95\linewidth]{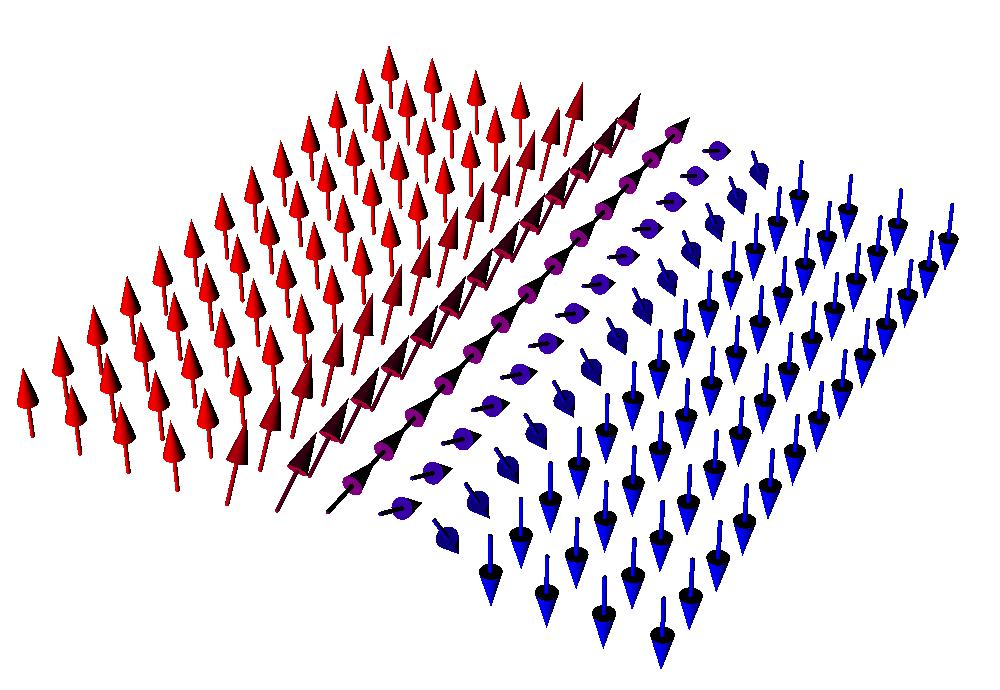} \\ (a)}
\end{minipage}
\hfill
\begin{minipage}[h]{0.24\linewidth}
\center{\includegraphics[width=0.95\linewidth]{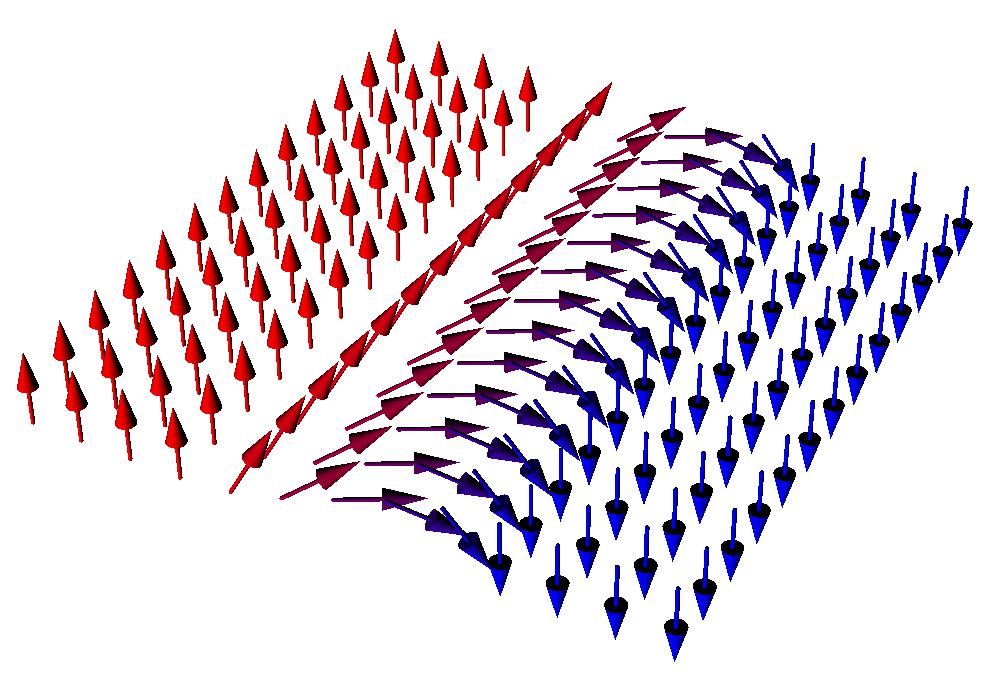} \\ (b)}
\end{minipage}
\hfill
\begin{minipage}[h]{0.24\linewidth}
\center{\includegraphics[width=0.95\linewidth]{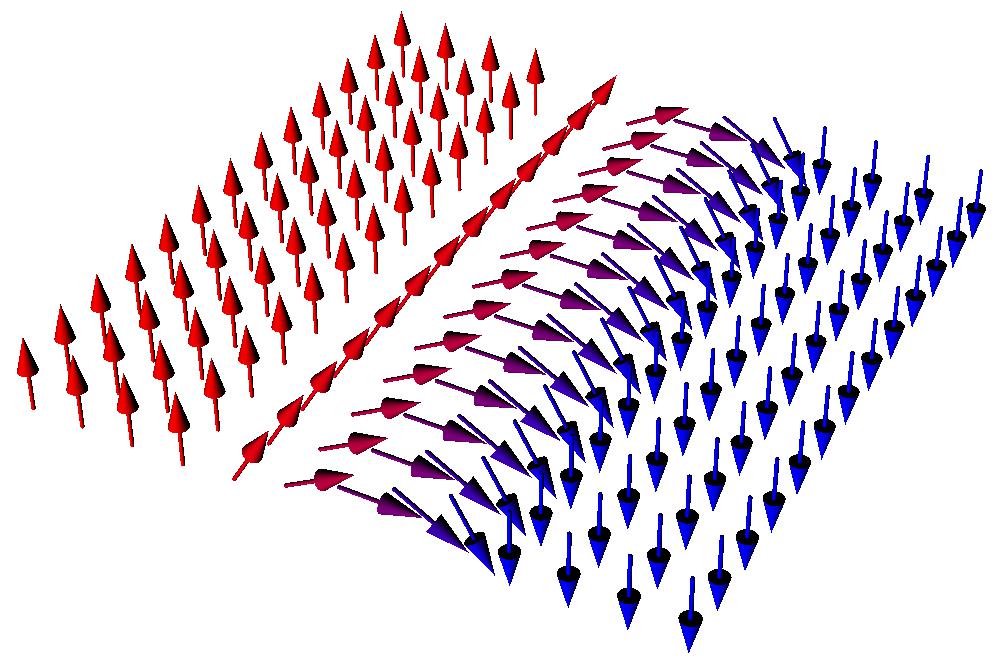} \\ (c)}
\end{minipage}
\hfill
\begin{minipage}[h]{0.24\linewidth}
\center{\includegraphics[width=0.95\linewidth]{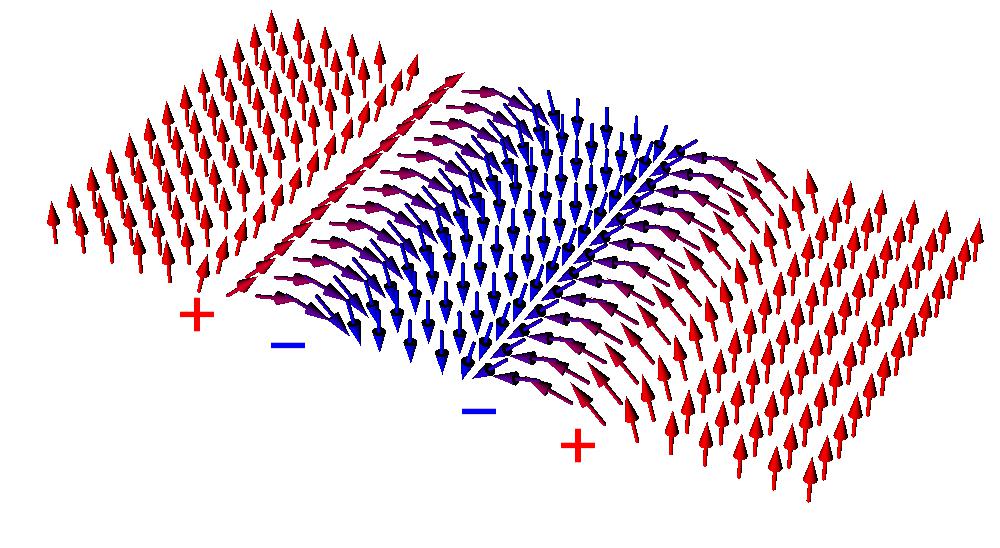} \\ (d)}
\end{minipage}
\caption{\label{Fig3} (a) -- Bloch, (b) -- canted, (c) --  N\'eel and (d) -- 360 degrees N\'eel domain walls with schematically shown magnetostatic charges distribution.}
\end{figure}

The following experimental observations may indicate the presence of a nonzero iDMI. The first one is existence of striped-type contrast on L-TEM images with the Fresnel contrast (Figs.~\ref{TITAN}b) in a some range of external fields (excitations of the objective lens). The micromagnetic nature of stripes is the following.  Domain walls in magnetic film with perpendicular magnetic anisotropy are Bloch-type (\ref{Fig3}a) in the absence of iDMI. Such distribution is favorable due to minimal magnetostatic energy. A weak iDMI tends to tilt the domain wall (\ref{Fig3}b) as iDMI minimizes the energy for magnetic distribution with nonzero magnetostatic charges $\left(\vec{\nabla} \cdot \vec{M} \right) \neq 0$. When the iDMI interaction is larger then a critical value~\cite{Meier}, the domain wall becomes pure N\'eel-type (\ref{Fig3}c). This value for our magnetic parameters is $\left| D_{cr} \right|=0.44 \mu_0 M_s^2 t/ \pi = 52.8$ $\mu J/m^2$, where $t$ is the film thickness. Sharp steps on hysteresis curve (point ``2'' at Fig.~\ref{Fig2}) as between 20-30 mT appears only for $\left| D \right| \gtrsim D_{cr}$.  The magnetization distribution at this point is a topological 360 degrees domain walls (Fig.~\ref{Fig3}d), which look like black-white stripes in Fresnel mode (Fig.~\ref{MicromagSim}b\&d). The interplay of magnetostatic and iDMI leads to the situation when two 180 degrees N\'eel domain walls are too close but they can’t annihilate (on the contrary to Bloch walls). The iDMI value for micromagnetic simulation can be selected easily from experimental data. On the one hand it should be larger than critical value $D_{cr}$, on the other hand a large iDMI (about $65-75 \mu J/m^2$) leads to formation of isolated N\'eel skyrmions instead of 360 degrees walls.

\begin{figure}
\begin{minipage}[h]{0.24\linewidth}
\center{\includegraphics[width=0.95\linewidth]{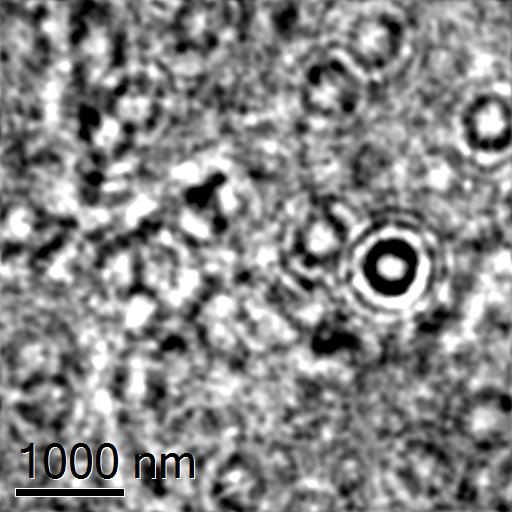} \\ (a)}
\end{minipage}
\hfill
\begin{minipage}[h]{0.24\linewidth}
\center{\includegraphics[width=0.95\linewidth]{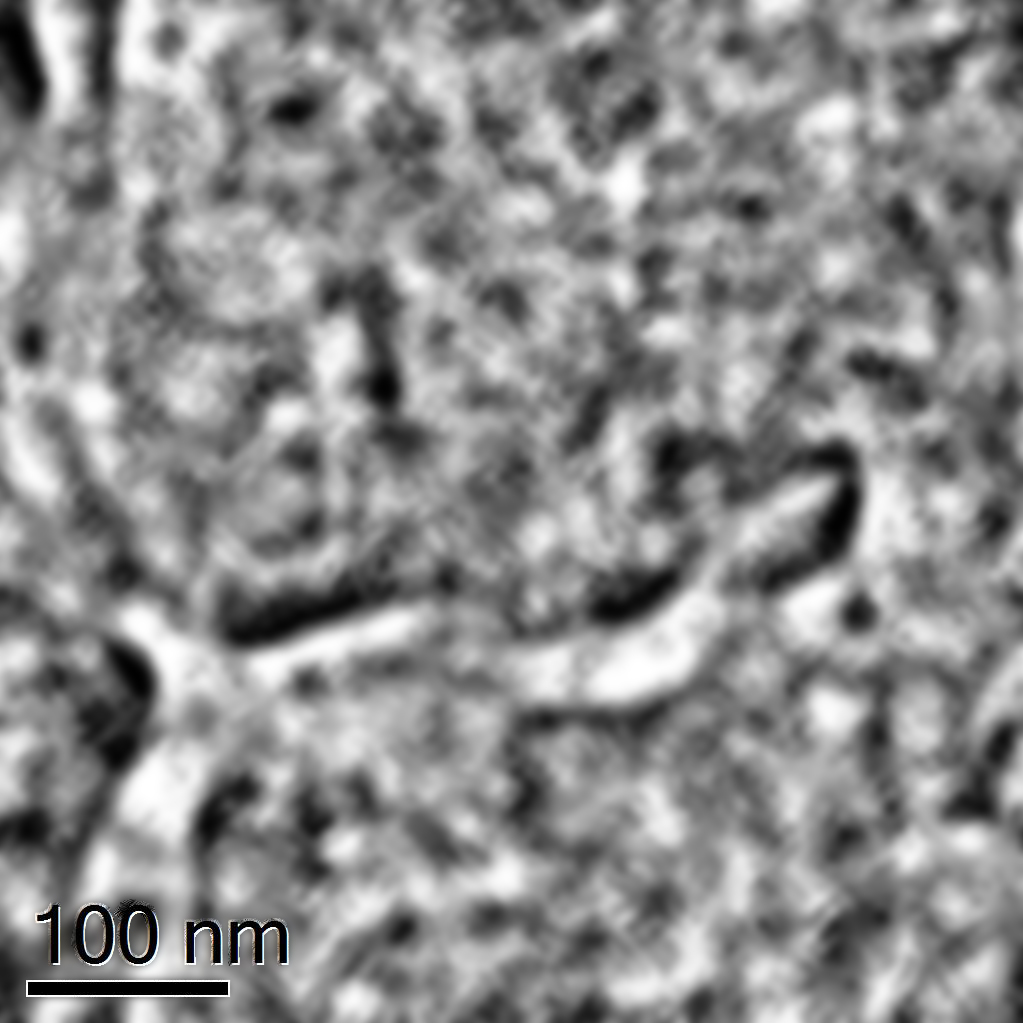} \\ (b)}
\end{minipage}
\hfill
\begin{minipage}[h]{0.24\linewidth}
\center{\includegraphics[width=0.95\linewidth]{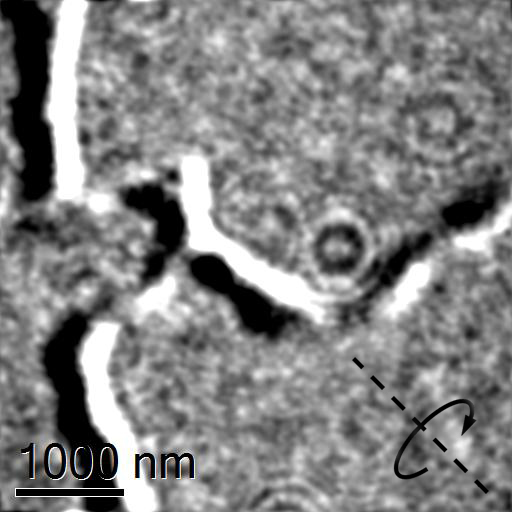} \\ (c)}
\end{minipage}
\hfill
\begin{minipage}[h]{0.24\linewidth}
\center{\includegraphics[width=0.95\linewidth]{StripesTilt} \\ (d)}
\end{minipage}
\caption{\label{360FresnelSim} Comparison of Fresnel and simulated contrast from 360 degrees N\'eel domain wall.}
\end{figure}

We also have observed weak Fresnel contrast at zero tilting angles. The explanation is the following. In case of pure N\'eel walls in homogeneous film there should be zero magnetic contrast at zero tilt~\cite{Walls01}. The analysis of Fresnel contrast with simulation of polycrystalline nature of magnetron sputtered films shows nonzero contrast for pure N\'eel walls at zero tilt conditions (Fig.~\ref{360FresnelSim}). The small contrast variations arise due to the magnetization tends to lie along wandering axis of the uniaxial magnetic anisotropy of nanocrystalline film structure. The tilting angle is 30 degrees along diagonal of the image. The origin of contrast on Fig.~\ref{360FresnelSim}a\&b is following. Wandering anisotropy axes leads to a weak canting of spins in domain wall from pure N\'eel configuration. That leads to a nonzero values for magnetization rotor and gives rise to local Fresnel contrast. Due to nanocrystalline origin of this contrast it is of the same order as the contrast from small magnetization cants of ``homogeneously'' magnetized areas. Thus the nanocrystalline structure of magnetron sputtered films can be the origin of weak magnetic contrasts in polycrystalline films~\cite{Walls05,Walls06}.

\section{Conclusion}
Summing up, we experimentally have investigated the crystal and magnetic structure of the Co/Pt multilayer films with perpendicular magnetic anisotropy. Basing upon some previous research~\cite{TEM2} we see the significant role of interlayers in formation of homochiral magnetic textures. In this paper we have shown the existense of interlayer contribution in iDMI even in case of ``symmetrical'' material structure of multilayer. It is surprising fact that the magnetron sputtered Co/Pt multilayers demonstrate the existence of a strong iDMI. Lorentz transmission electron microscopy and MOKE measurements analysed with micromagnetic simulations confirm the iDMI value above critical, which makes it possible to stabilize chiral distribution of magnetization such as N\'eel domain walls in perpendicular magnetic anisotropy media. Also a careful analysis of L-TEM micrographs shows the presence of a weak magnetic Fresnel contrast due to the influence of the wandering axis of uniaxial magnetic anisotropy in nanocrystalline grains. Such interesting behaviour of Co/Pt multilayers (without any third material, such as Pd, Ta, W, Ir etc.) can simplify the technology of production thin magnetic films, hosting homochiral N\'eel skyrmions at room temperatures.

\begin{acknowledgments}
The work was supported by Russian Foundation for Basic Research Grant \# 18-02-00827. The equipment of Physics and Technology of Micro- and Nanostructures (Institute for Physics of Microstructures, Russian Academy of Sciences) was used. The equipment of CIC nanoGUNE BRTA (SAN Sebastian, Spain) is used for performing LTEM measurements. The authors are grateful to A.A. Fraerman and M.V. Sapozhnikov for fruitful discussion; M.Yu. Mikhailovskii and P.A. Yunin for sample preparation; K.V. Tatarskaya for help with the manuscript preparation.
\end{acknowledgments}

\bibliography{WormsCoPt}

\end{document}